\documentclass[runningheads]{llncs}
\usepackage{graphicx}
\usepackage{blindtext}
\usepackage{color}
\begin{document}
\title{Joint Management and Analysis of Textual Documents and Tabular Data within the AUDAL Data Lake}
\titlerunning{Data Management and Analysis within the AUDAL Data Lake}
%
\author{Pegdwendé N. Sawadogo\inst{1} \and
Jérôme Darmont\inst{1} \and 
Camille Noûs\inst{2}} 
\authorrunning{P.N. Sawadogo et al.}
%
\institute{Université de Lyon, Lyon 2, UR ERIC \\
5 avenue Pierre Mendès France, F69676 Bron Cedex, France
\email{\{pegdwende.sawadogo,jerome.darmont\}@univ-lyon2.fr}\\
\and
Université de Lyon, Lyon 2, Laboratoire Cogitamus \\
\email{camille.nous@cogitamus.fr}\\
}

\maketitle              
\begin{abstract}
In 2010, the concept of data lake emerged  as an alternative to data warehouses for big data management. 
Data lakes follow a schema-on-read approach to provide rich and flexible analyses. 
However, although trendy in both the industry and academia, the concept of data lake is still maturing, and there are still few methodological approaches to data lake design. 
Thus, we introduce  a new approach to design a data lake and propose an extensive metadata system to activate richer features than those usually supported in data lake approaches.  
We implement our approach in the AUDAL data lake, where we jointly exploit both textual documents and tabular data, in contrast with structured and/or semi-structured data typically processed in data lakes from the literature. Furthermore, we also innovate by leveraging metadata to activate both data retrieval and content analysis, including Text-OLAP and SQL querying. 
Finally, we show the feasibility of our approach using a real-word use case on the one hand, and a benchmark on the other hand. 

\end{abstract}
\section{Introduction}

Over the past two decades, we have witnessed a tremendous growth of the amount of data produced in the world. These so-called big data come from diverse sources and in various formats, from social media, open data, sensor data, the Internet of things, etc. Big data induce great opportunities for organizations to get valuable insights through analytics. 
However, this presupposes storing and organizing data in an effective manner, which involves great challenges. 

Thus, the concept of data lake was proposed 
 to tackle the challenges related to the variety and velocity characteristics of big data~\cite{Dixon2010}. 
A data lake can be defined as a very large data storage, management and analysis system that handles any data format. Data lakes use a schema-on-read approach, i.e., no schema is fixed until data are analyzed~\cite{Fang2015}, which provides more flexibility and richer analyses than traditional storage systems such as data warehouses, which are based on a schema-on-write approach~\cite{Khine2017}. 
Yet, in the absence of a fixed schema, analyses in a data lake heavily depend on metadata~\cite{Hai2018}. 
Thus, metadata management plays a vital role. 

Although quite popular in both the industry and academia, the concept of data lake is still maturing. 
Thence, there is a lack of methodological proposals for data lakes implementations for certain use cases. Existing works on data lakes indeed mostly focus on structured and/or semi-structured data \cite{Hai2016,Halevy2016,Maccioni2018,Mehmood2019}, with little research on managing unstructured data. Yet, unstructured data represent up to 80\% of the data available to organizations~\cite{Diamantini2018}. Therefore, managing texts, images or videos in a data lake is an open research issue. 

Furthermore, most of data lake proposals (about 75\%) refer to Apache Hadoop for data storage
~\cite{Russom2017}. However, using Hadoop requires technical human resources that small and medium-sized enterprises (SMEs) may not have. Thence, alternatives are needed.
Last but not least, data lake usage is commonly reserved to data scientists~ \cite{Fang2015,Khine2017,Madera2016}. Yet, business users represent a valuable expertise while analyzing data. Consequently, opening data lakes to such users is also a challenge to address.

 To meet these issues, we contribute to the literature on data lakes through a new approach to build and exploit a data lake. We implement our approach in AUDAL (the AURA-PMI\footnote{AURA-PMI is a multidisciplinary project in Management and Computer Sciences, aiming at studying the digital transformation, servicization and business model mutation of industrial SMEs in the French Auvergne-Rhône-Alpes (AURA) Region.} Data Lake). AUDAL exploits an extensive metadata system to activate richer features than common data lake proposals.
 More concretely, our contribution is threefold.
 First, we introduce a new methodological approach to integrate both structured (tabular)  and unstructured (textual) data in a lake. Our proposal opens a wider range of analyses than common data lake proposals, which goes from data retrieval to data content analysis. Second, AUDAL also innovates through an architecture leading to an ``inclusive data lake'', i.e, usable by data scientists as well as business users. Third, we propose an alternative to Hadoop for data and metadata storage in data lakes. 
 
 The remainder of this paper is organized as follows. 
 In Section~\ref{sec:metadata.management}, we focus on our metadata management approach. 
 In Section~\ref{sec:archi.analyses}, we detail AUDAL's architecture and the analyses it allows. In  Section~\ref{sec:eval}, we demonstrate the feasibility of our approach through performance measures. In Section~\ref{sec:related.works}, we review and discuss the related works from the literature. Finally, in Section~\ref{sec:conclusion}, we conclude the paper and hint at future research.

\section{Metadata Management in AUDAL}
\label{sec:metadata.management}

The most critical component in a data lake is presumably the metadata management system. In the absence of a fixed schema, accessing and analyzing the lake's data indeed depend on metadata~\cite{Hai2016,Maccioni2018,Suriarachchi2016}.
Thence,  we particularly focus in this section on how metadata are managed in AUDAL. 

First and foremost, let us precise what we consider as metadata. We adopt the definition: ``structured information that describes, explains, locates, or otherwise makes it easier to retrieve, use, or manage information resources'' \cite{Visengeriyeva2020}. This definition highlights that metadata are not limited to simple atomic data descriptions, but may be more complex.

AUDAL's metadata management system is based on MEDAL~\cite{Sawadogo2019B}, a metadata model for data lakes. We adopt MEDAL because it is extensive  enough to allow both data exploration and data content analysis by business users. 
In line with MEDAL, our metadata system implements data polymorphism, i.e., the simultaneous management of multiple raw and/or preprocessed representations of the same data~\cite{Sawadogo2019B}. Our motivation is that different analyses may require the same data, but in various, specific formats. Thus, pregenerating several formats for data would lead to 
readily available and faster analyses~\cite{Armbrust2021,Leclercq2018}. 


 Still in line with MEDAL, we use the  term ``object'' as our lower-granularity data item, i.e., either a tabular or textual document. We also exploit three types of metadata that are detailed in  
the following sections.  Section~\ref{subsec:intra.manag} is dedicated to \textit{intra-object metadata} management;  Section~\ref{subsec:inter.manag}  focuses on \textit{inter-object metadata} management; and  Section~\ref{subsec:global.manag} details \textit{global metadata} management. 
.


\subsection{Intra-object Metadata}
\label{subsec:intra.manag}

\subsubsection{Definition and Generation}
Intra-object metadata are atomic or more complex information associated with a specific object. 
We classify them in two categories. 

\textit{Metadata properties} are information that describe an object. They often take the form of simple key-value pairs, e.g., author name, file path, creation date, etc. 
However, they may sometimes be more complex. Particularly, the description of the columns of a table can be viewed as a complex form of metadata properties. 

Metadata properties are mostly provided by the file system. However, especially when dealing with textual documents, we use Apache Tika
~\cite{ApacheTika}
to automatically extract metadata such as the author, language, creation timestamp, mime-type and even the program used to edit the document. 

\textit{Refined representations} are more complex. 
When an object is transformed, the result may be considered as both data and metadata. This is in line with the definition we adopt for metadata, since such transformed data make easier the use of the original object. 
In AUDAL, refined representations of textual documents are either bag-of-word vectors~\cite{Pu2007} or document embedding vectors~\cite{Le2014}. 
Bag-of-words can easily be aggregated to extract top keywords from a set of documents. However, they do not suit distance calculation, due to their high dimensionality. By contrast, embedding vectors do not bear this disadvantage, while allowing the extraction of top keywords. 
Refined representations of tabular data are plain and simply relational tables. 
Eventually, let us note that AUDAL's metadata system may be extended with additional types of refined representations, if needed. 

To generate bag-of-word representations, we perform for each document a classical process:  tokenizing, stopword removal, lemmatization and finally word count. 
To generate embedding representations, we project documents in an embedding space with the help of the Doc2Vec model~\cite{Le2014}. Each document is thus transformed into a reduced vector of only a few tens of coordinates.
Eventually, we use a custom process to generate refined representations from tables. Each tabular document is read in a Python dataframe and then transformed into a relational table.  

\subsubsection{Modeling and Storage}
Still using MEDAL~\cite{Sawadogo2019B}, we follow a graph approach to model the interactions between data and metadata.  
Therefore, AUDAL's metadata system is centered on Neo4J~\cite{Neo4J}.
We exploit four types of nodes to manage intra-object metadata. 

 \textit{Object} nodes  represent raw objects. They contain atomic metadata, i.e., metadata properties, in particular the path to the raw file. 
As Neo4J does not support non-atomic data inside nodes, we define  \textit{Column} nodes to store column descriptions. \textit{Column} nodes are thus associated to \textit{Object} nodes only in the case of tabular documents.   

Each \textit{Object} node is also associated with \textit{Refined} nodes that reference refined representations stored in other DBMSs. Refined representations of textual documents, i.e., embedding and bag-of-word vectors, are indeed stored in MongoDB~\cite{MongoDB}. Similarly, refined representations of tabular documents  are stored in the form of SQLite 
tables~\cite{SQLite}.
\textit{Refined} nodes stored in Neo4J actually contain references to their storage location. 



\begin{figure*}[hbt] 
\includegraphics[width=\textwidth]{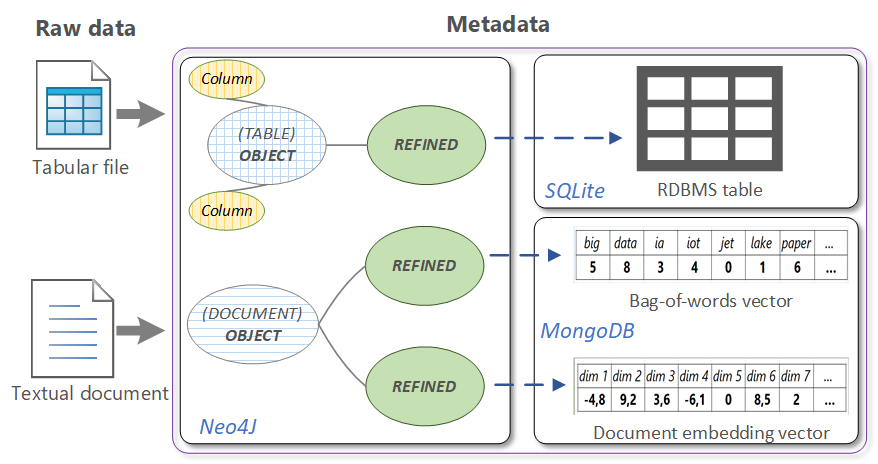}
\caption{Organization of intra-object metadata in AUDAL}
 \label{fig:intra}
\end{figure*}

Figure~\ref{fig:intra} illustrates the organization of intra-object metadata.

\subsection{Inter-object Metadata}
\label{subsec:inter.manag}

\subsubsection{Definition and Generation} Inter-object metadata are information that reflect relationships between objects. 
We manage two types of inter-object metadata.

\begin{figure*}[hbt] 
\includegraphics[width=\textwidth]{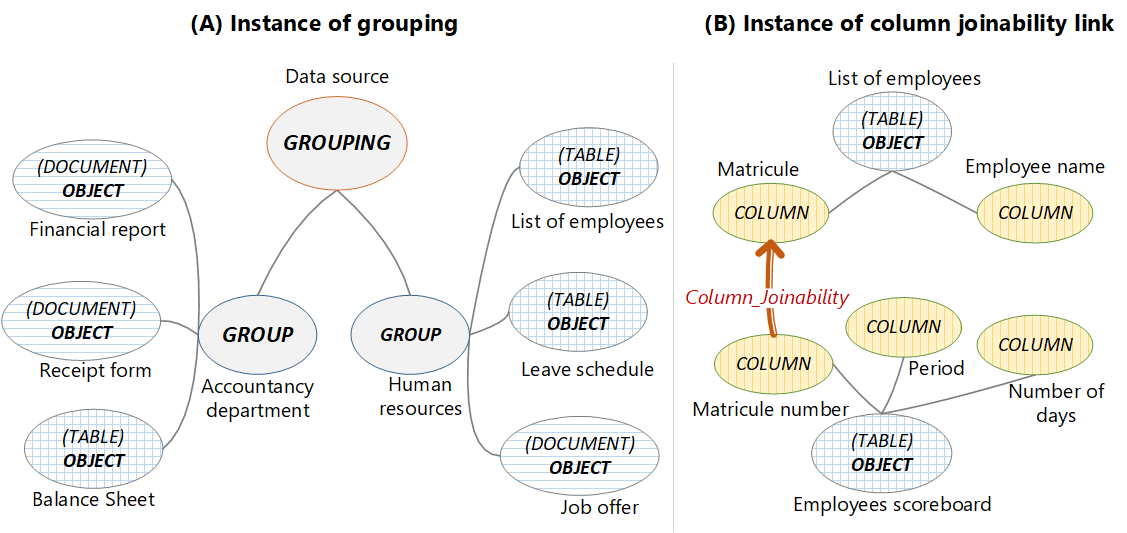}
\caption{Organization of inter-object metadata in AUDAL}
 \label{fig:inter}
\end{figure*}

\textit{Data groupings} are organized tag systems that allow to categorize objects into groups, i.e., collections. Each data grouping induces several groups, i.e., collections. Then, data retrieval can be achieved through simple intersections and/or unions of groups. Data groupings are particularly interesting as they are not data type-dependent. For example, a grouping based on data source can serve to retrieve tabular data as well as textual documents, indistinctly (Figure~\ref{fig:inter}A). 

Data groupings
are usually generated on the basis of categorical properties. Starting from the property of interest, we first identify possible groups. Then, each object is associated to the group it belongs to.  

\textit{Similarity links}  are information on relatedness between objects. These metadata are obtained applying a suitable similarity measure between couple of textual documents. In our case, we use the cosine similarity that is classically used in information retrieval to compare document vector-space representations~\cite{Allan2000}. As the number of potential links increases exponentially with the number of documents, we simply retain the links of each document to its ten closest.  
When dealing with tabular data, 
we use  primary key/foreign key relationships to link columns and thus connect tables. We deduce  primary key/foreign key relationships from raw data with the help of the PowerPivot method~\cite{Chen2014}, which is casually used in structured data lakes~\cite{Fernandez2018}. 

\subsubsection{Modeling and Storage}
To set up data groupings, we introduce two types of nodes in AUDAL's metadata catalogue: \textit{Grouping} and \textit{Group}. A \textit{Grouping} node represents the root of a partition of objects. Each \textit{Grouping} node is associated with several \textit{Group} nodes that represent the resulting parts of such a partition. For example, a partition on data source could lead to a \textit{Group} node for ``accounting department'', another for ``human resources'' and so on~(Figure~\ref{fig:inter}A). \textit{Object} nodes are then associated with \textit{Group} nodes with respect to the group they belong to. 
A data grouping organization may thus be seen as a three-layer tree graph where the root node represents the grouping instance, intermediate nodes groups, and leaf nodes objects.

More simply, 
similarity measures in AUDAL are 
edges linking nodes. Such edges potentially carry information that indicates the strength of the link, how it was measured, its orientation, etc. 
More concretely, textual similarity is represented by edges of type \textit{Document\_Similarity} between Neo4J \textit{Object} nodes. We model tabular data similarity with \textit{Coulumn\_Joinability} edges between \textit{Column} nodes to connect primary key/foreign key column pairs that appear to be joinable. Figure~\ref{fig:inter}B depicts an instance of \textit{Column\_Joinability} edge that connects two tables through columns.


\subsection{Global Metadata}
\label{subsec:global.manag}

\subsubsection{Definition and Generation}
Global metadata are data structures that are built and continuously enriched to facilitate and optimize analyses in the lake. 
AUDAL includes two types of global metadata.

\textit{Semantic resources} are knowledge bases  (thesauri, dictionaries, etc.) that help improve both metadata generation and data retrieval. Dictionaries allow filtering on specific terms and building  vector representations of documents. Similarly, AUDAL uses a thesaurus to automatically expand term-based queries with synonyms.  
Such semantic resources are ingested and enriched by lake users.

\textit{Indexes} are also exploited in AUDAL. An inverted index is notably a data structure that establishes a correspondence between keywords and objects from the lake. Such an index is particularly needed to support and, above all, speed-up term-based queries.  
There are two indexes in AUDAL: $document\_index$ and  $table\_index$. The first handles the entire content of each textual document, while the latter  collects all string values in tabular documents 




\subsubsection{Modeling and Storage}
As global metadata are not directly linked to objects, we do not focus on their modeling, but on their storage, instead. 
In AUDAL, we manage indexes with ElasticSearch~\cite{ElasticSearch}, an open-source indexing service that enforces scalability.  
We define in ElasticSearch an alias to allow simultaneous querying on the two indexes. 
Eventually, we store semantic resources, i.e., thesauri and dictionaries, 
in a MongoDB collection. Each is thus a MongoDB document that can be updated and queried.

\section{AUDAL's Architecture and Analysis Features}
\label{sec:archi.analyses}

In this section, we highlight how AUDAL's components are organized~(Section~\ref{subsec:archi}) and the range of possible analyses (Section~\ref{subsec:analyses}).

\subsection{AUDAL Architecture}
\label{subsec:archi}
\begin{figure*}[hbt] 
\centering
\includegraphics[width=10cm]{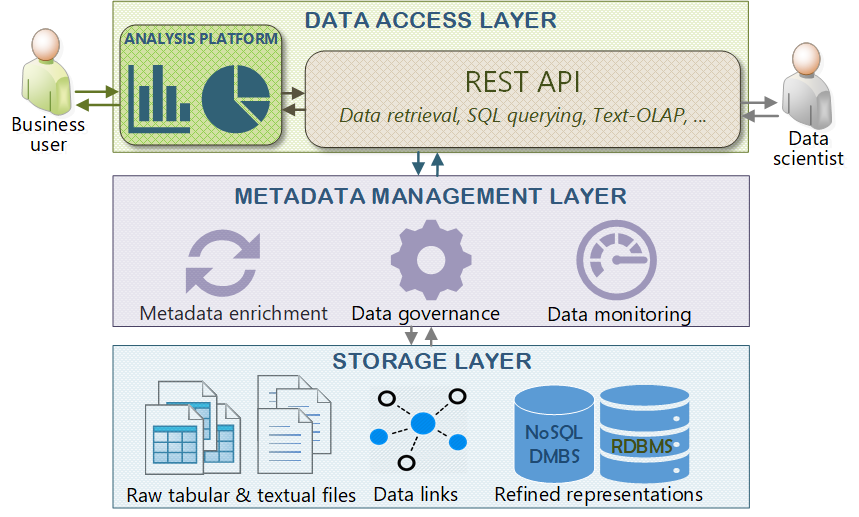}
\caption{Architecture of AUDAL}
 \label{fig:archi}
\end{figure*}

AUDAL's functional architecture is made of three main layers: a storage layer, a metadata management layer and a data querying layer~(Figure~\ref{fig:archi}).

The \textbf{storage layer} is in charge of storing raw and processed data, as well as metadata, through a combination of storage systems, each adapted to a specific storage need. In AUDAL, we use a simple file system for raw data storage, a graph DBMS to store links across data entities, a document-oriented DBMS to store refined representations and a relational DBMS for table storage.  

The \textbf{metadata management layer} is made of a set of processes dedicated to data polymorphism management. More concretely, this layer is in charge of generating metadata, notably refined representations from raw data, as well as links. It allows future analyses and avoids a data swamp, i.e., a data lake whose data cannot be accessed~\cite{Suriarachchi2016}. The data swamp  syndrome is indeed often caused by a lack of efficient metadata management system.

Finally, the \textbf{data querying layer} is an interface that consumes data from the lake. 
Its main component is a representational state transfer application programming interface (REST API) from which raw data and some ready-to-use analyses are accessible to data scientists. 
However, a REST API is not accessible to business users who, unlike data scientists, do not have enough skills to transform raw data into useful information on their own. In addition, business users are not familiar with API querying. Thence, we also provide a graphical analysis platform for them  in AUDAL. 
This platform features the same functions as the REST API, but in a graphical way. 
Thus, each type of user can access the lake with respect to its needs, which makes AUDAL ``inclusive'', unlike the common vision of data lakes that excludes business users~\cite{Fang2015,Khine2017,Madera2016}.


Overall, AUDAL's architecture looks a lot like a multistore system, i.e., a collection of heterogeneous storage systems with a uniform query language~\cite{Leclercq2018}. AUDAL indeed offers a single REST API to query data and metadata across different systems (Neo4J, 
MongoDB, ElasticSearch, 
and SQLite). 
However, AUDAL also features an extensive metadata management layer that goes beyond what  multistore systems do, i.e., multistores handle only intra-object metadata.  

\subsection{AUDAL's Analysis Features}
\label{subsec:analyses}

\subsubsection{Data Retrieval}
Data retrieval consists in filtering data from the lake. 
The techniques we propose for data retrieval are suitable for both textual and tabular documents. 

\textit{Term-based querying} allows to filter data with respect to a set of keywords 
 It includes a fuzzy search feature that allows to expand queries with syntactically similar terms. 

\textit{Navigation} exploits groupings, i.e., organized sets of tags that allow data filtering by intersecting several groups. 
For example, we can retrieve documents edited by a given author on a specific year, who is associated with a department via the intersection of three groups, e.g., ``Scott'', 2010 and ``Human resources''.

\textit{Finding related data} consists retrieving the objects that are the closest of a given object. 
Closeness is obtained from similarity links. For example, in the case of tabular data, we use \textit{Column\_Joinability} links.   

\subsubsection{Document Content Analysis}
Content analyses provide insights from one or several objects, while taking their intrinsic characteristics into account. The techniques we propose are  specific to each data type. 
In the case of textual documents, AUDAL allows OLAP-like analyses~\cite{Codd1993}.
Groupings may indeed serve as dimensions and thus allow data filtering in multiple manners.
Thus, the lake's data can quickly and intuitively be reduced to a subset by intersecting groups, which is comparable to OLAP Slice \& Dice operations.

Once documents are filtered, they can be aggregated to obtain valuable insights.  
Aggregated results can be compared across different subsets of documents using suitable visualizations. 


\textit{Top keywords} summarize documents through a list of most frequent keywords, by aggregating a bag-of-word representation of documents. Thanks to the principle of data polymorphism, different top keyword extraction strategies can coexist. For instance, one can be based on a predefined vocabulary, while an other is based on a free vocabulary. We graphically display top keywords using bar charts or word clouds.  
    
\textit{Scoring} numerically evaluates the relatedness of a set of documents to a set of query terms with the help of a scoring algorithm that takes into account, amongst others, the appearances of query terms in each document. Due to the wide number of documents, the scores per document may not be readable. Thence, we propose instead an aggregated score per group.  

\textit{Highlights} display text snippets where a set of terms appear. In other words, it can be viewed as a document summary centered on given terms. This is also commonly called a concordance.

\textit{Group comparison} exploits embedding representations to show together 
groups of documents using a similar vocabulary. 
This is done in two steps. First, we average the embedding vectors of all documents per group. Then, 
we exploit the resulting mean embedding vectors to extract group likeness using KMeans clustering~\cite{Jain2010} or principal component analysis (PCA)~\cite{Wold1987}. KMeans analysis identifies strongly similar groups into a user-defined number of clusters, while PCA provides a simple two-dimensional visualization where the proximity between groups reflects their similarity. 

\subsubsection{Tabular Content Analysis}
We propose several ways to 
analyze tabular data. 

\textit{SQL querying} helps users extract or join tabular data. 
SQL queries actually run on the refined representations of tabular data. As such refined representations are in the form of classical relational tables, 
all SQL features are supported, including joins and aggregations. 

\textit{Column correlation} evaluates the links between a couple of table columns. We use a suitable statistical measure  with respect to columns types. For example, a Jaccard similarity measure can serve to compare categorical columns, while the Kolmogorov-Smirnov statistic 
is suitable for numerical columns~\cite{Bogatu2020}.

\textit{Tuple comparison} consists in running a KMeans clustering or a PCA on a set of tuples, 
 by taking only numeric values into account. Tuples to compare are extracted through a SQL query, potentially including joins and/or 
 aggregations. 

\section{Quantitative Assessment of AUDAL}
\label{sec:eval}
The goal of the experiments we propose in this section is to show the feasibility and adaptability of our approach. For this purpose, we implement AUDAL with two different datasets. 
AUDAL's source code is available online\footnote{https://github.com/Pegdwende44/AUDAL}.
\subsection{Datasets and Query Workload}
The first dataset we use comes from the AURA-PMI project. 
It is composed of 8,122 textual documents and 6 tabular documents, 
for a total size of 6.2~GB.
As the AURA-PMI dataset is quite small, we also create an artificial dataset 
by extracting 50,000 scientific articles from the French open archive HAL. 
To these textual documents, we add 5,000 tabular documents coming from an existing benchmark~\cite{Nargesian2018B}, for a total volume of 62.7~GB. 

To compare how AUDAL works on our two datasets, we define 
a set of 15 queries that reflect AUDAL's main features (Table~\ref{tab:queries}). Then, we measure the response time of our workload to assess whether our approach is realistic. In Table~\ref{tab:queries}, the terms document, table and object refer to textual document, relational table and one or the other indistinctly, respectively.

\begin{table}[tb]
\caption{Query workload}
\begin{center}
\begin{tabular}{|c|l|}

     \hline
     \multicolumn{2}{|c|}{\textbf{\textit{Data retrieval queries}}} \\
    \hline
    1 & Retrieve documents written in English and edited in December \\
    \hline
    2 & Retrieve objects (tables or documents)  containing the terms ``big'' and ``data''   \\
    \hline
    3 & Retrieve objects with terms ``big'', ``data'', ``document'' and ``article'' \\
    \hline
    4 & Retrieve 3 tables, joinable to any table. \\
    \hline
    5 & 
    Retrieve 5 most  similar documents to a given document\\ 
    \hline
    \hline
     \multicolumn{2}{|c|}{\textbf{\textit{Textual content analysis}}} \\
    \hline
     6 & Calculate document scores w.r.t. the terms ``big'', ``data'', ``article''\\
        & and ``document'' \\
    \hline
    7 & Extract a concordance from documents using the terms ``data'' and ``ai'' \\
    \hline
    
    8 & Extract a concordance from documents using the terms ``data'', ``ai'' ``article'' \\ 
        & and ``paper''  \\
    \hline
    9 & Find top 10 keywords from all documents   \\
    \hline
    10 & Run a 3-cluster KMeans clustering on documents grouped by month \\
    \hline
    11 & Run a PCA analysis on documents grouped by month. \\
     \hline
     \hline
     \multicolumn{2}{|c|}{\textbf{\textit{Tabular content analysis}}} \\
    
    \hline
    12 &  Run a join query between two tables \\
    \hline
    13 & Run a join query between two tables while averaging all numerical values \\
        & and aggregating by any categorical column. \\
    \hline
    14 & Run a 3-cluster KMeans clustering on the result of \textit{query 12} \\
    \hline
     15 & Run a PCA on the result of \textit{query 12}. \\
    \hline

\end{tabular}
\end{center}
\label{tab:queries}
\end{table}


\subsection{Experimental Setup and Results}
Both instances of AUDAL are implemented on a cluster of three VMware virtual machines (VMs). The first VM  has a 7-core Intel-Xeon 2.20~GHz processor and 24~GB of RAM. It runs the API. Both other VMs have a mono-core Intel-Xeon 2.20~GHz processor and 24~GB of RAM. Each hosts a Neo4J instance, an ElasticSearch instance and a MongoDB instance to store AUDAL's metadata. 
The execution times we report in Table~\ref{tab:results1} are the average of ten runs of each query, expressed in milliseconds. 
 \begin{table}[hbt]
\begin{minipage}{.5\textwidth}
 
\caption{Query response time (ms)}
\begin{center}

\begin{tabular}{|c|c|c|}

     \hline
      \textbf{Query} & \textbf{AURA-PMI} & \textbf{Artificial} \\
       & \textbf{dataset} & \textbf{dataset} \\
     \hline
     \hline
     \multicolumn{3}{|c|}{\textbf{\textit{Data retrieval queries}}} \\
    \hline
    Query 1 & 194 &  653 \\
    \hline
    Query 2 & 108 &   207 \\
    \hline
    Query 3 & 143 & 305 \\
    \hline
    Query 4 & 59 & 81 \\
    \hline
    Query 5 & 51 & 79 \\ 
    \hline
    \hline
     \multicolumn{3}{|c|}{\textbf{\textit{Textual content analysis}}} \\
    \hline
     Query 6 & 85 & 117 \\
    \hline
   Query 7 & 169 & 198 \\
    \hline
    
   Query 8 & 62 &  92 \\
    \hline
   Query 9 & 4,629&   188,199\\
    \hline
   Query 10 & 1,930 & 26,969
 \\
    \hline
   Query 11 & 1,961& 26,871 \\
     \hline
     \hline
     \multicolumn{3}{|c|}{\textbf{\textit{Tabular content analysis}}} \\
    
    \hline
   Query 12 & 71 & 37 \\
    \hline
   Query 13 & 61 & 12 \\
   
    \hline
   Query 14 & 174 & 144 \\
    \hline
    Query 15 & 670 & 520 \\
    \hline

\end{tabular}
\end{center}
\label{tab:results1}

\end{minipage}
\begin{minipage}{.5\textwidth}
  \begin{center}

\caption{Raw data vs. metadata size (GB)}
\begin{tabular}{|c|c|c|}

     \hline
      \textbf{System} & \textbf{AURA-PMI} & \textbf{Artificial} \\
       & \textbf{dataset} & \textbf{dataset} \\
     \hline
     \hline
  \multicolumn{3}{|c|}{\textbf{\textit{Raw data}}} \\
    \hline
  
    - & \textbf{6.2} & \textbf{62.7} \\
      \hline
      \hline
      \multicolumn{3}{|c|}{\textbf{\textit{Metadata}}} \\
    \hline
    Neo4J & 0.9 & 2.0 \\
    \hline
    SQLite & 0.003 &  1.7 \\
    \hline
    MongoDB & 0.28 & 3.4\\
    \hline
    ElasticSearch & 1.6 & 27.6\\
    \hline
    \textbf{Total }& \textbf{2.8} &\textbf{34.7} \\ 
    \hline
    
    \hline

\end{tabular}
\end{center}
\label{tab:results2}
\end{minipage}
\end{table}

Our experimental results show that AUDAL does support almost all its query and analysis features in a reasonable time. 
We also see that AUDAL scales quite well with respect to data volume. All data retrieval and tabular content analyses indeed run very fast 
 on both the AURA-PMI dataset (174~ms on average) and the larger, artificial dataset  (183~ms on average). 
Admittedly, half of textual content queries, i.e., queries \#9, \#10 and \#11, take longer to complete: 5, 2 and 2 seconds on average, respectively, on the AURA-PMI dataset; and 188, 27 and 27 seconds on average, respectively, on the artificial dataset. 
However, we note that without our approach, such tasks would be definitely impossible for business users. Moreover, the situation can certainly be improved  by increasing CPU resources. 
Thus, we consider our results promising. 

However, AUDAL's features are achieved at the cost of an extensive metadata system. Table~\ref{tab:results2} indeed shows that the size of metadata represents up to half of raw data. Yet we deem this acceptable given the benefits. Moreover, it is acknowledged that metadata can be larger than the original data, especially in the context of data lakes, where metadata are so important~\cite{Hellerstein2017}.



\section{Related Works}
\label{sec:related.works}
The research we present in this paper relates to many systems from the data lake literature. Some of them address data retrieval issues, while others mostly focus on data content analysis. We discuss them with respect of our structured and unstructured data context.

\subsection{Data Retrieval from Data Lakes}
A great part of the literature considers data lakes as a playground dedicated to data scientists. Related research focuses on data retrieval, since content analyses are assigned to expert users.      
We identify three main approaches for data retrieval in data lakes, namely navigation, finding related data and term-based search. 
A first retrieval-by-navigation model exploits 
tags to easily and quickly find the target object~\cite{Nargesian2018B}. A similar approach is implemented in several data lakes~\cite{Bagozi2019,Halevy2016,Mehmood2019}. However, all these models are set in the context of structured data only. 

A second data retrieval approach exploits data relatedness, i.e., finding a significant similarity between objects or their components~\cite{Bogatu2020}. Several techniques help 
detect relatedness between tabular data through column joinability and unionability~\cite{Bogatu2020,Farrugia2016,Fernandez2018,Maccioni2018}. 
To the best of our knowledge, only one proposal~\cite{Diamantini2018} is relevant to  unstructured data.

Finally, 
term-based querying is particularly useful for textual data. Thus, in previous work, we used an indexing system to allow textual documents data retrieval~\cite{Sawadogo2019A}. This technique, i.e., inverted indexes,  is also implemented with structured data in Google's data lake~\cite{Halevy2016} and CoreKG~\cite{Beheshti2018}.

\subsection{Data Content Analysis from Data Lakes}
An alternative vision of data lakes considers that business users, i.e., not data scientists, can also consume data from a lake. Thus, content querying is required and methods must be used to ease the users' work.
In the structured data world, fuzzy SQL querying can be used in data lakes~\cite{Mrozek2018}. Similarly, a custom query rewriting system is exploited to analyse data from the Constance lake~\cite{Hai2018}. There is also a way to personalize table querying by taking user profile into account~\cite{Bagozi2019}.
Although very few, some approaches propose content analysis for semi-structured \cite{Hai2016} and unstructured data~\cite{Sawadogo2019A}. 
The latter exploits text and graph mining techniques to enable document aggregation. 

\subsection{Discussion}
As stated above, most data lake approaches focus either on data retrieval or data content analyses. Therefore, they present a partial vision of data lakes, in our opinion. In contrast, there exists a system that frees itself from this cleavage~\cite{Bagozi2019}. However,  it does not support unstructured data. 
More generally, unstructured data are very rarely supported in data lakes. Our own CODAL data lake~\cite{Sawadogo2019A} does manage textual documents management, but only textual documents. It is therefore limited.
In contrast, AUDAL goes beyond these limitations by featuring both data retrieval as well as content analyses. In addition, AUDAL supports both tabular documents and, above all, textual documents whose inclusion in data lakes still challenging. 

\section{Conclusion and Future Works}
\label{sec:conclusion}
In this paper, we present AUDAL, 
presumably the first methodological approach to manage both textual and tabular documents in a data lake. 
AUDAL includes an extensive metadata system to allow querying and analyzing the data lake 
and supports more features than state-of-the-art data lake implementations. In terms of queries, AUDAL indeed supports both data retrieval and data content analyses, including Text-OLAP and SQL querying. 
Moreover, AUDAL also allows the exploitation of a data lake not only by data scientists, but also by business users. All these makes AUDAL an ``inclusive'' data lake. 

In our near-future research, we plan a deeper validation of AUDAL on two aspects. First, we will work on that complexity and time cost of metadata generation algorithms. Second, we will study how AUDAL's analysis interface is useful to and usable by business users, e.g., using the widely used SUS (System Usability Scale) protocol~\cite{Brooke1996}.
Another perspective is data lineage tracking to allow AUDAL support version management. This is particularly important for tabular documents that are often merged or altered. Such a lineage could be implemented by extending AUDAL's refined representations. 
Finally, we envisage to include more unstructured data types into a lake, i.e., images, videos and/or sounds, and manage their particular metadata for retrieval and analysis.

\section*{Acknowledgments}
P.N. Sawadogo's PhD is funded by the Auvergne-Rhône-Alpes Region through the AURA-PMI project.

%
%
%
 \bibliographystyle{splncs04}
 \bibliography{biblio}

\end{document}